\documentclass[aps,pre,twocolumn,nofootinbib,floatfix,showpacs]{revtex4}  

\usepackage{graphicx} 
\newcommand {\be}{\begin{equation}}
\newcommand {\ee}{\end{equation}}
\newcommand {\ba}{\begin{eqnarray}}
\newcommand {\ea}{\end{eqnarray}}

\begin{document}

\title{Efficient time series detection of the strong stochasticity threshold in Fermi-Pasta-Ulam oscillator lattices}

\author{M. Romero-Bastida}
\email{mromerob@ipn.mx}
\author{Alan Yoshio Reyes-Mart\'\i nez}
\affiliation{SEPI ESIME-Culhuac\'an, Instituto Polit\'ecnico Nacional, Av. Santa Ana No. 1000, Col. San Francisco Culhuac\'an, Delegaci\'on Coyoacan, Distrito Federal 04430, Mexico}

\date{\today}

\begin{abstract}
In this work we study the possibility of detecting the so-called strong stochasticity threshold, i.e. the transition between weak and strong chaos as the energy density of the system is increased, in anharmonic oscillator chains by means of the 0-1 test for chaos. We compare the result of the aforementioned methodology with the scaling behavior of the largest Lyapunov exponent computed by means of tangent space dynamics, that has so far been the most reliable method available to detect the strong stochasticity threshold. We find that indeed the 0-1 test can perform the detection in the range of energy density values studied. Furthermore, we determined that conventional nonlinear time series analysis methods fail to properly compute the largest Lyapounov exponent even for very large data sets, whereas the computational effort of the 0-1 test remains the same in the whole range of values of the energy density considered with moderate size time series. Therefore, our results show that, for a qualitative probing of phase space, the 0-1 test can be an effective tool if its limitations are properly taken into account.
\end{abstract}


\maketitle

\section{Introduction\label{sec:Intro}}

Deterministic chaos, defined as a dynamical regime with sensitive dependence on initial conditions, can be readily quantified by the largest Lyapunov exponent (LLE) $\lambda$, which is an averaged exponential growth rate of linear perturbations to the considered motion~\cite{Ott}. The LLE is easily computed in numerical simulation, but it is difficult to obtain in experiment, where usually there is no knowledge of the nonlinear equations that govern the time evolution of the system. Hence, a number of alternative chaos-detection techniques that rely on the temporal evolution of a reduced set of variables have been developed over the years with varying degrees of success~\cite{Kantz}. 

The 0-1 test~\cite{Test1}, which uses as input the time series of an observed variable, was designed to distinguish regular behavior, if 0 is the output, from chaotic dynamics, if the output is 1, in deterministic dynamical systems. It has been successfully applied to both numerically obtained~\cite{Test2,Barrow,Dawes,Litak1,Litak2} as well as to simple experimental data~\cite{SIADS}. However, in a previous work~\cite{TestMRB} we have presented evidence that, when applied to time series stemming from nonhomogeneous, i.e dissimilar mass, many-degrees-of-freedom systems, the test has serious limitations as an efficient chaos detector for a large system size $N$. On the other hand, for small homogeneous systems, but in a very low energy regime (indistinguishable of regular dynamics for very large time scales and termed \emph{weak chaos}), the test misclassifies the provided signal as regular, even though the LLE has a vanishingly small non-negative value.

Of the foregoing results, the latter is by far the most important to be taken into account when applying the 0-1 test, since the homogeneity condition can be readily controlled in a simulation or in an experiment, and systems with a small number of homogeneous components are indeed physically relevant, as will be discussed below. The analysis of Ref.~\cite{TestMRB} makes it plausible to infer that, for homogeneous mass systems composed of a small number of particles, it can be generically expected that, in a low energy (or temperature) regime, the 0-1 test will render a result close to zero notwithstanding the actual dynamical regime of the considered system. Now, if a microscopic model is available, in principle the LLE provides a way to know, with no uncertainty, the type of dynamical regime wherein the system is, since the dynamical equations give information of all the existing range of scales for arbitrarily long time intervals. However, there are evident practical problems with the computation of such quantity, since the aforementioned conditions, besides being unattainable when dealing with experimental data, may also result of no physical interest whatsoever, since it is now clear that the LLE is not completely satisfactory for a proper characterization of the many faces of complexity and predictability of many-degrees-of-freedom systems; see~\cite{Cencini} and references therein, where some examples are discussed showing that systems with different dynamics can give similar results when analyzed from a time series point of view. From this perspective, what can be validly ascertained is that, if the result of the 0-1 test is $0$ for a signal obtained from a simulation or experiment of a homogeneous system in a very low energy or temperature regime, the dynamics can be regarded as regular for the considered time series length, regardless of what the outcome of the test would be with a larger (and maybe inaccessible) data set.

The precise knowledge of the actual dynamical regime may not be a crucial piece of information if the main interest is to probe other phase-space structural characteristics of many-degrees-of-freedom systems that go beyond the simple distinction between order and chaos. In particular, the so called Fermi-Pasta-Ulam (FPU) oscillator chain presents a transition from weak to strong chaos as the energy density $\epsilon\equiv E/N$ of the lattice increases beyond a threshold value $\epsilon_{_T}$ known as the strong stochasticity threshold (SST)~\cite{Pettini,Pettini2}. The LLE $\lambda(\epsilon)$ exhibits a change in its scaling behavior precisely at the value $\epsilon_{_T}$ of the SST. If a complete knowledge of the dynamics is available, even an analytical estimate of the LLE energy dependence for both low and high energies can be readily obtained~\cite{Casetti96}. However, with only limited information (such as the time series of the position and/or momentum of a single oscillator in an actual experiment) the computation of the LLE becomes problematic. Therefore, it would be interesting to explore the possibility that the 0-1 test could detect the aforementioned transition, regardless the misclassification (which, in principle, can only be entirely avoided in the infinite time limit) of the actual dynamical regime for very low $\epsilon$ values obtained by means of this method. This result could be specially relevant for the case of experimental data with no \emph{a priori} information about the microscopic dynamics, but with precise structural information available that can rule out the limitations addressed in Ref.~\cite{TestMRB}.

This paper is organized as follows. In Sec.\ \ref{sec:t01} we present the relevant detail of the 0-1 test and in Sec.\ \ref{sec:Model} we describe the employed model and the relevant details of its numerical integration. Sec.\ \ref{sec:Results} presents the results of the detection of the SST by means of the 0-1 test as well as a comparison with results stemming from standard nonlinear time series analysis methods. In Sec.\ \ref{sec:Disc} we discuss the previous results and present our conclusions.

\section{The 0-1 test for chaos\label{sec:t01}}

We follow the implementation of the 0-1 test reported in~\cite{Test2}: See Ref.~\cite{Test4} for the relevant mathematical results that justify the version herein employed. The test takes as input a finite data set $\{\phi(t_{\alpha})\}_{\alpha=1}^{\cal N}$ sampled at discrete times $t_{\alpha}\equiv\alpha\tau$, with sampling time $\tau$. For a given $c\in\Re$ we construct the transformed series $\xi(t_{\alpha})=\sum_{j=1}^{\alpha}\phi(t_j)\cos(jc)$, $\alpha=1,2,3,\ldots$ Now, from the mean square displacement defined as
\be
M(t_{\alpha})=\lim_{{\cal N}\rightarrow\infty}{1\over{\cal N}-\alpha}\sum_{j=1}^{{\cal N}-\alpha}\left[\xi(t_{j+\alpha}) - \xi(t_{\alpha})\right]^2~\label{MSD}
\ee
we compute numerically the asymptotic growth rate $K=\lim_{\alpha\rightarrow\infty}(\log M(t_{\alpha}))/\log t_{\alpha}$ by performing a least square fit of $\log M(t_{\alpha})$ versus $\log t_{\alpha}$ in the range $1\le\alpha\le{\cal N}_1$ for a choice of ${\cal N}_1$ such that $1\ll{\cal N}_1 \ll{\cal N}$ and ${\cal N}_1={\cal N}/10$. We compute $K$ for $100$ random values of $c$ and the final $K$ value is taken as the median of the computed set. Then $K\approx0$ stands for regular dynamics and $K\approx1$ for chaotic dynamics.

\section{the model and its numerical simulation\label{sec:Model}}

The Hamiltonian of the one-dimensional $N$ coupled FPU lattice from which the employed time series were obtained reads as
\be
H=\sum_{i=1}^N\left[{p_i^2\over2m_i} + {1\over2}(q_{i+1}-q_i)^2 + {1\over4}(q_{i+1}-q_i)^4\right] \label{HFPU},
\ee
where $\{m_i,q_i,p_i\}_{i=1}^N$ are the dimensionless mass, displacement, and momentum of the $i$ oscillator, respectively. Periodic boundary conditions are assumed. Equilibrium displacements $\{q_i(0)=0\}$ and momenta $\{p_i(0)\}$ distributed according to a Maxwell-Boltzmann distribution consistent with the desired energy density $\epsilon$ are taken as initial conditions. The homogeneity of the system is assured by taking a unit mass value $m_i=1$ for all oscillators. Next, the $2N$ first-order Hamilton equations of motion were integrated using a symmetrical version of the velocity Verlet algorithm (VVA)~\cite{Tuck}. With the adopted value $\Delta t=0.01$ of the time step a faithful representation of a Hamiltonian flow and a driftless value of total energy $E$ for the studied time scales is ensured.

\section{results\label{sec:Results}}

\subsection{Comparison with tangent space methods}

\begin{figure}
\includegraphics[width=0.90\linewidth,angle=0.0]{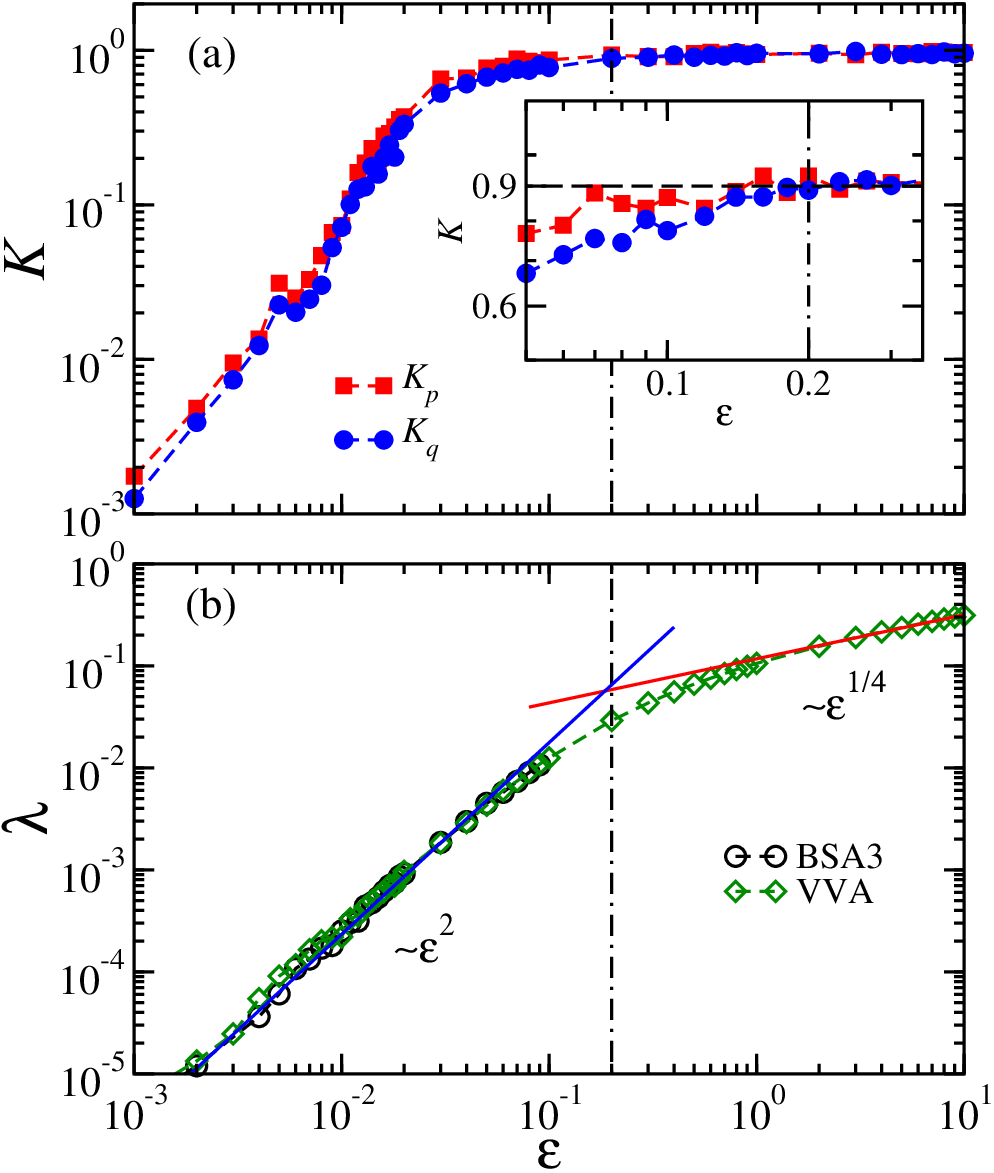}
\caption{(Color online) (a) Asymptotic growth rate $K$ vs energy density $\epsilon$ computed from the position (circles) and momentum (squares) time series of a single oscillator of a FPU chain with $N=32$. The inset presents in more detail the transition region where the $K$ values computed from both time series become almost indistinguishable. The horizontal dashed line indicates the value $0.9$ to which both $K_q$ and $K_p$ converge for large $\epsilon$ values. (b) LLE vs $\epsilon$ obtained by means of the standard method. Diamonds correspond to the LLE computed using the VVA of Ref.~\cite{Tuck}, whereas circles refer to data obtained by means of the BSA3 of Ref.~\cite{Casetti}. Continuous lines in (b) refer to the scaling laws valid for low and high $\epsilon$ values, whereas vertical dot-dashed lines in both (a) and (b) indicate the approximate value $\epsilon_{_T}$ corresponding to the SST.}
\label{fig:kly}
\end{figure}

For each considered $\epsilon$ value the position and momentum time series of a single oscillator of a FPU lattice with $N=32$ were recorded with a sampling time of $\tau=1$ (that corresponds to 100 time steps for $\Delta t=0.01$) with a total time series length of ${\cal N}=10^5$. The 0-1 test was applied to the so obtained data sets and the corresponding asymptotic growth rates $K_q$ (position) and $K_p$ (momentum) for each $\epsilon$ was computed. The results are displayed in Fig.~\ref{fig:kly}(a). As can be readily appreciated, $K$ grows steadily and gradually reaches a $\sim1$ value as the energy density changes from low to high values. The transition between the aforementioned behaviors occurs at $\epsilon\in[0.1,0.2]$. A further refinement is obtained if one observes that $K_q$ and $K_p$ have different (but albeit very close) values up to the threshold $\epsilon_{_{01}}\approx0.2$, where the difference $K_q-K_p$ becomes minimal and $K_q,K_p\approx0.9$ consistently hereafter, as seen in the inset of the same figure. The behavior of $K$ vs $\epsilon$ indicates a transition from a very ordered dynamical regime to a strongly chaotic one, with a very precise threshold value $\epsilon_{_{01}}$ separating both regimes. This capability of the 0-1 test to probe detailed dynamical information has not been previously acknowledged. The exact quantification of the degree of exponential divergence, which also conveys information of the aforementioned transition, is obtained by means of the LLE. These results for LLE $\lambda(\epsilon)$, computed by means of the standard method~\cite{benettin80,shimada79}, are presented in Fig.~\ref{fig:kly}(b). The SST $\epsilon_{_T}$, obtained from the intersection of the scaling laws (whose theoretical explanation is considered in Ref.~\cite{Casetti96}) valid for low and high $\epsilon$ values, has a remarkable agreement with $\epsilon_{_{01}}$ obtained by means of the 0-1 test and is consistent with other estimates~\cite{FPUHLM}. Since it is known that a precise computation of the LLE is difficult in the low energy density regime~\cite{Data}, we have performed some further computations with the more sophisticated third order bilateral symplectic algorithm (BSA3)~\cite{Casetti}. An excellent agreement with the results obtained by means of the VVA can be readily appreciated in that same figure.

\subsection{Results of phase-space reconstruction methods}

\begin{figure}
\includegraphics[width=0.90\linewidth,angle=0.0]{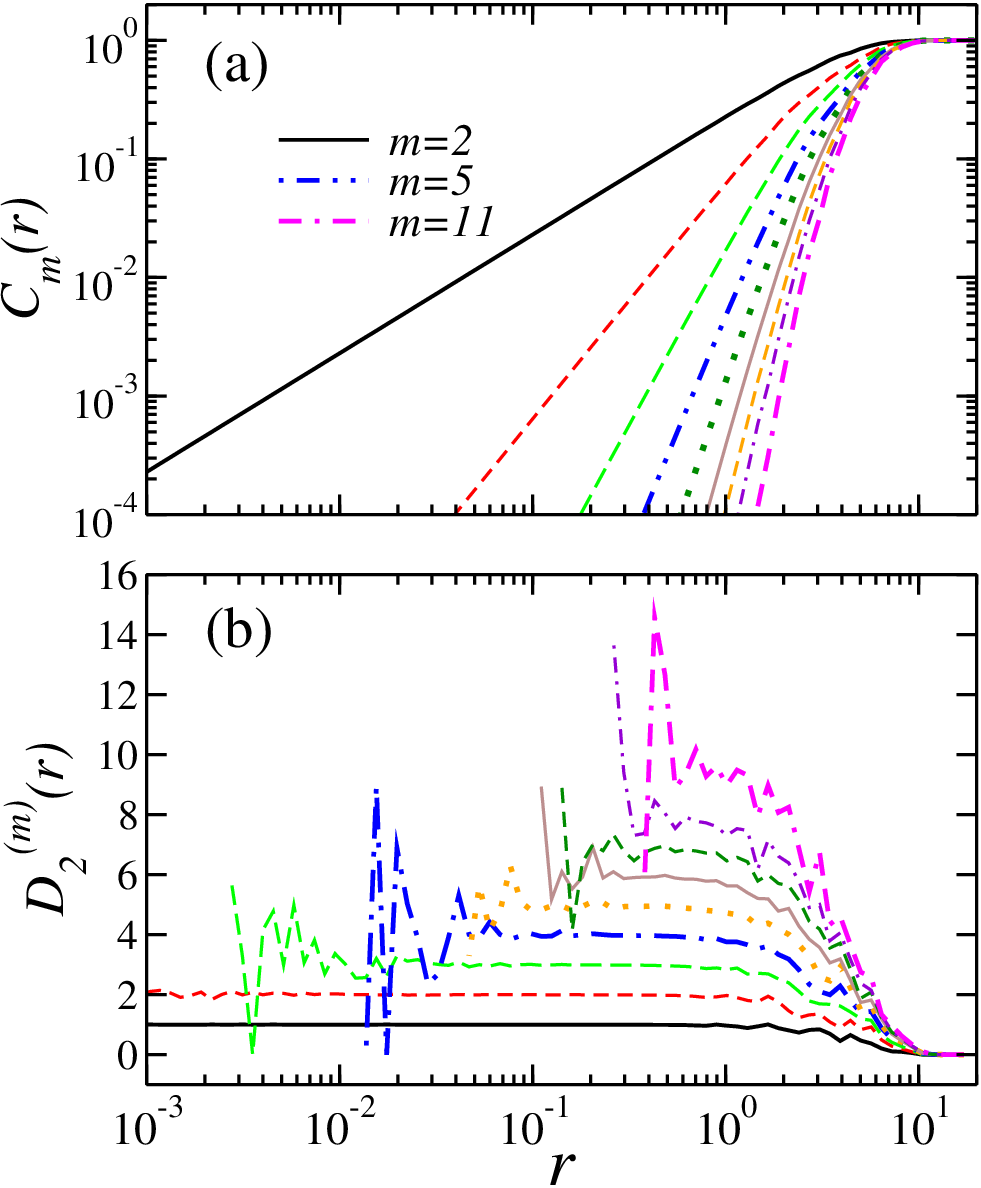}
\caption{(Color online) (a) Correlation integral $C_m(r)$ and (b) dimension $D_2^{(m)}(r)$ vs resolution $r$ computed from the position time series of a single oscillator of a FPU lattice with $N=32$ and $\epsilon=10$, with embedding dimension $m$ increasing from left to right in (a) and from bottom to top in (b).}
\label{fig:cd2}
\end{figure}

Notice that the knowledge of the tangent space dynamics is not usually available when studying experimental time series, since it can only be obtained from an explicit model. Therefore, the LLE has to be obtained from different approximation techniques such as phase space reconstruction methods, successfully applied to dissipative systems~\cite{Kantz}. In order to corroborate if the LLE of the Hamiltonian model herein studied can indeed be computed by means of these techniques we proceed to determine an embedding dimension $m$ that can approximate the underlying dimensionality $2N$ of the considered dynamical system~\cite{GP}. In Fig.~\ref{fig:cd2} we present the results for both the correlation integral $C_m(r)$ and dimension $D_2^{(m)}(r)$ for the position time series of a single oscillator of the FPU lattice for a range of different spatial resolutions $r$; see Ref.~\cite{Kantz} for further details. As can be appreciated in Fig.~\ref{fig:cd2}(a), it is not possible to obtain a $m$-independent scaling range for the correlation integral. Furthermore, from Fig.~\ref{fig:cd2}(b) it is clear that the larger $m$ the smaller is the scaling range for the correlation dimension, with remarkable statistical fluctuations at lower resolutions. Recalling that on large length scales the data cannot be distinguished from random noise~\cite{Olbrich97,Cencini}, it becomes evident that not even the dynamical character of the provided signal can be inferred from this methodology. The problem can be traced back to the insufficient length of the employed time series. However, and since already for time series of $10^6$ data points the computation time becomes prohibitively large, it is clear that the computation of an adequate embedding dimension is impractical for a $N=32$ FPU lattice. Thus, it is unfeasible to infer the dimensionality of this system from the available data by means of phase reconstruction methods

\begin{figure}
\includegraphics[width=0.90\linewidth,angle=0.0]{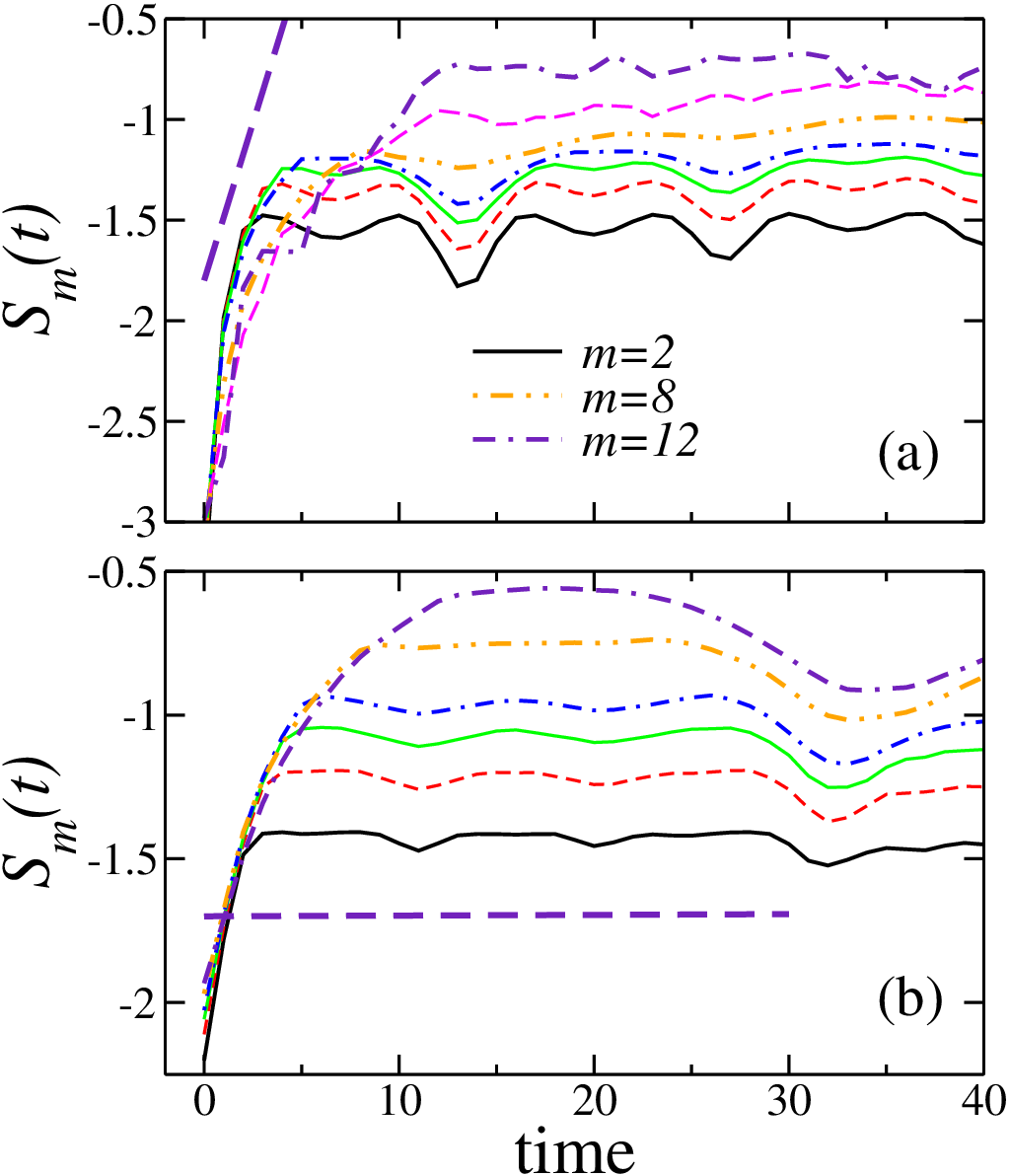}
\caption{(Color online) (a) Average local distance $S_m(t)$ vs time for the data set used in Fig.~\ref{fig:cd2}, with embedding dimension increasing from bottom to top. Slope of dashed line corresponds to the LLE value displayed in Fig.~\ref{fig:kly}(a). (b) Same as (a), but for a time series corresponding to $\epsilon=0.01$.}
\label{fig:lyts}
\end{figure}

The foregoing results impose severe restrictions to the computation of the LLE from time series, since it is known that the number of data points required to estimate the LLE is about the square of that needed to estimate the embedding dimension~\cite{ER92}. Nevertheless we explore the extent of the aforementioned limitation by applying the algorithm proposed by Kantz in Ref.~\cite{Kantz94}, which does not depend explicitly on the knowledge of the correct embedding dimension, to our system. The results are displayed in Figs.~\ref{fig:lyts}(a) and (b), where $S_m(t)$, which is the temporal average of a suitable measure to the distances between all neighboring trajectories to a reference trajectory inside a $r$-neighborhood for a given embedding dimension $m$, is plotted for low and high $\epsilon$ values. It is clear that no temporal interval can be clearly identified, for all resolutions $r$ studied, where $S_m(t)$ exhibits a linear increase with identical slope for any embedding dimension value larger than $m=2$. Therefore, no estimation of the LLE can be obtained from the employed time series, even in the strongly chaotic regime. It is important to mention that the data employed in the aforementioned calculation are exactly the same ones supplied to the 0-1 test to compute the corresponding $K$ value displayed in Fig.~\ref{fig:kly} and that these results are virtually unaltered when time series ten times larger are considered. Therefore, at least for this particular problem, the 0-1 test dramatically outperforms traditional phase space reconstruction methods, which seemingly cannot cope with the dimensionality of the underlying dynamical system from where the time series originated. Our results are more conclusive than those in Ref.~\cite{Test2}, where the LLE of the dissipative eight-dimensional Lorenz 96 system could actually be computed with the method of Rosenstein \emph{et al.}~\cite{Rosenstein93} and compared to the corresponding $K$ values rendered by the 0-1 test. 

\section{discussion and conclusions\label{sec:Disc}}

The preceding results clearly indicate that a quantitative estimate of the $\epsilon_{_T}$ value corresponding to the SST can be efficiently obtained by means of the 0-1 test for chaos. The importance of this estimation stems from the fact that the SST has been detected in one-dimensional lattices with a $\phi^4$ interaction potential~\cite{Pettini}, as well as with Toda, smoothed Coulomb, and Lennard-Jones potentials~\cite{Yoshimura}; in an isotropic Heisenberg spin chain~\cite{Constantoudis}; in a mean field XY chain~\cite{Firpo} and in a coupled rotator chain which displays two thresholds separating two regions of weak chaos (occurring at low and high energies) from an intermediate region of strong chaos~\cite{XY}. It has also been detected in two- and three-dimensional lattices with two-well $\phi^4$~\cite{Caiani1,Caiani2} and XY Heisenberg interactions~\cite{Butera,Caiani3}. In particular, for a three-dimensional system consisting of a small number of ions confined in a Penning optical trap and forming a so-called microplasma the dependence of the LLE on the energy is very similar to that displayed by the aforementioned models where the detection of the SST has been performed~\cite{Microplasmas1}. Furthermore, a very detailed analysis of the behavior of the LLE as a function, not only of the energy but also of other parameters that control the geometry of the trap, is available~\cite{Microplasmas1}.

At this point it is important to emphasize that the detection of the SST in the FPU model (or of a similar phenomenology in microplasmas, as already mentioned) does not require the precise knowledge of the LLE value: any observable (such as the asymptotic growth rate $K$) that shares the same dependence on the energy density with the LLE will be equally useful as well. From this observation, and recalling that the trajectory of each individual ion in an optical trap can be observed and tracked experimentally, it is then highly feasible that much of the already known results obtained by the LLE for confined ions could be reproduced with less effort by means of the 0-1 test and corroborated by analogous calculations performed on experimental data. Furthermore, the maximum ion number $N=40$ employed in Ref.~\cite{Microplasmas1} is of the same order than the oscillator number $N=32$ in the present study; therefore, it can be reasonably expected that even the number of data points needed to characterize the microscopic dynamics of this microplasma by means of the 0-1 test for chaos could be of the same order than that used to obtain the asymptotic growth rate $K$ in Fig.~\ref{fig:kly} for the FPU lattice. Since, as already mentioned, it is a difficult task to compute the LLE from experimental records, this simple and efficient technique could complement, and even become a viable alternative when a rapid qualitative analysis is called for, to the approximate methods to the tangent space dynamics recently applied to characterize the microscopic dynamics of confined ions~\cite{Microplasmas2}.

We wish to end our discussion by mentioning that ``cuspy" patterns of the energy dependence of the LLE, in contrast to the mild transition displayed by the FPU model, show up in the presence of a thermodynamic phase transition, as in the case of the mean-field XY model and the two-dimensional lattice with $\phi^4$ interactions previously mentioned. Now, it has been proposed that the appearance of singularities in the thermodynamic observables could be the effect of a suitable topological transition in subspaces of configuration space~\cite{PettiniBook}. It could be interesting to explore the possibility that information about such detailed dynamical features could be probed by means of the herein presented methodology. In this context, another interesting possibility would be to test the feasibility of obtaining the abrupt transitions just mentioned from time series stemming from variables different from the thermodynamic observables, which present singularities at phase transitions and are thus difficult to measure experimentally. 

\smallskip
\begin{acknowledgments}
The author wishes to thank Juan M.~L\'opez, Diego Paz\'o, M.~C.~Nu\~nez-Santiago and Monica Sofia Romero-Nu\~nez for their valuable comments and suggestions. Financial support from CONACyT, M\'exico is also acknowledged.
\end{acknowledgments}




\begin{thebibliography}{10}

\bibitem{Ott}
E. Ott, {\em Chaos in Dynamical Systems} (Cambridge University Press,
  Cambridge, 1993).

\bibitem{Kantz}
H. Kantz and T. Schreiber, {\em Nonlinear Time Series Analysis} (Cambridge
  University Press, Cambridge, 1997).

\bibitem{Test1}
G.~A. Gottwald and I. Melbourne, Proc. R. Soc. London A {\bf 460},  603
  (2004).

\bibitem{Test2}
G.~A. Gottwald and I. Melbourne, Physica D {\bf 212},  100  (2005).

\bibitem{Barrow}
J.~D. Barrow and J. Levin, arXiv:nlin.CD/0303070 (unpublished).

\bibitem{Dawes}
J.~H.~P. Dawes and M.~C. Freeland (unpublished).

\bibitem{Litak1}
G.~Litak, A.~Syta, and M.~Wiercigroch, Chaos Solitons Fractals {\bf 40},  2095
  (2009).

\bibitem{Litak2}
G.~Litak, A.~Syta, M.~Budharja, and L.~M. Saha, Chaos Solitons Fractals {\bf
  42},  1511  (2009).

\bibitem{SIADS}
I.~Falconer, G.~A. Gottwald, I. Melbourne, and K. Wormnes, SIAM J. Appl. Dyn.
  Syst. {\bf 6},  395  (2007).

\bibitem{TestMRB}
M.~Romero-Bastida, M.~A. Olivares-Robles, and E.~Braun, J. Phys. A: Math.
  Theor. {\bf 42},  495102  (2009).

\bibitem{Cencini}
M.~Cencini, M.~Falcioni, E.~Olbrich, H.~Kantz, and A.~Vulpiani, Phys. Rev. E {\bf 62}, 427 (2000);
See also M.~Cencini, F. Cecconi, and A.~Vulpiani, {\em Chaos: From Simple Models to Complex Systems}
(World Scientific, Singapore, 2010).

\bibitem{Pettini}
M. Pettini and M. Landolfi, Phys. Rev. A {\bf 41},  768  (1990).

\bibitem{Pettini2}
M. Pettini and M. Cerruti-Sola, Phys. Rev. A {\bf 44},  975  (1991).

\bibitem{Casetti96} L.~Casetti, C.~Clementi, and M.~Pettini, Phys. Rev. E {\bf 54}, 5969 (1996).

\bibitem{Test4}
G.~A. Gottwald and I. Melbourne, Nonlinearity {\bf 22},  1367  (2009).

\bibitem{Tuck}
M. Tuckerman, B.~J. Berne, and G.~J. Martyna, J. Chem. Phys. {\bf 97},  1990
  (1992).

\bibitem{benettin80}
G. Benettin, L. Galgani, A. Giorgilli, and J.-M. Strelcyn, Meccanica {\bf 15},
  9  (1980).

\bibitem{shimada79}
I. Shimada and T. Nagashima, Prog. Theor. Phys. {\bf 61},  1605  (1979).

\bibitem{FPUHLM}
H.-L. Yang and G. Radons, Phys. Rev. E {\bf 73},  066201  (2006).

\bibitem{Data} Whereas for large $\epsilon$ values $15\times10^6$ time steps and reorthonormalizing each 100 steps was sufficient to obtain a reliable LLE estimate, in the opposite limit $9\times10^{11}$ times steps, with a reorthonormalization each 10000 steps, were necessary to obtain a similar precision level.

\bibitem{Casetti}
L. Casetti, Phys. Scr. {\bf 51},  29  (1995).

\bibitem{GP}
P. Grassberger and I. Procaccia, Phys. Rev. Lett. {\bf 50},  346  (1983).

\bibitem{Olbrich97}
E. Olbricht and H. Kantz, Physics Letters A {\bf 232},  63  (1997).

\bibitem{ER92}
J.-P. Eckmann and D. Ruelle, Physica D {\bf 56},  185  (1992).

\bibitem{Kantz94}
H. Kantz, Physics Letters A {\bf 185},  77  (1992).

\bibitem{Rosenstein93}
M.~T. Rosenstein, J.~J. Collins, and C.~J.~D. Luca, Physica D {\bf 65},  117
  (1993).

\bibitem{Yoshimura}
K. Yoshimura, Physica D {\bf 104},  148  (1997).

\bibitem{Constantoudis}
V. Constantoudis and N. Theodorakopoulos, Phys. Rev. E {\bf 55},  7612  (1997).

\bibitem{Firpo}
M.-C. Firpo, Phys. Rev. E {\bf 57},  6599  (1998).

\bibitem{XY}
L. Casetti, C. Clementi, and M. Pettini, Phys. Rev. E {\bf 54},  5969  (1996).

\bibitem{Caiani1}
L. Caiani, L. Casetti, and M. Pettini, J. Phys. A: Math. Gen. {\bf 31},  3357
  (1998).

\bibitem{Caiani2}
L. Caiani {\it et~al.}, Phys. Rev. E {\bf 57},  3886  (1998).

\bibitem{Butera}
P. Butera and G. Caravati, Phys. Rev. A {\bf 36},  962  (1987).

\bibitem{Caiani3}
L. Caiani, L. Casetti, C. Clementi, and M. Pettini, Phys. Rev. Lett. {\bf 79},
  4361  (1997).

\bibitem{Microplasmas1}
P. Gaspard, Phys. Rev. E {\bf 68},  056209  (2003).

\bibitem{Microplasmas2}
C. Antonopoulos, V. Basios, and T. Bountis, Phys. Rev. E {\bf 81},  016211
  (2010).

\bibitem{PettiniBook} M. Pettini, {\em Geometry and Topology in Hamiltonian Dynamics and Statistical Mechanics}
 IAM Series No. 33 (Springer, New York, 2007).

\end{thebibliography}
\end{document}